\documentclass[aps,prb,twocolumn,showpacs,amsmath,amssymb]{revtex4-1}
\usepackage{graphicx}
\usepackage{color}
\usepackage[colorlinks=true]{hyperref}

\newcommand{\tr}{\mathrm{tr}}
\newcommand{\calL}{\mathcal{L}}
\newcommand{\calT}{\mathcal{T}}
\newcommand{\bgamma}{\boldsymbol{\gamma}}
\newcommand{\bnabla}{\boldsymbol{\nabla}}
\newcommand{\bp}{\boldsymbol{p}}
\newcommand{\bx}{\boldsymbol{x}}
\newcommand{\bE}{\boldsymbol{E}}
\newcommand{\bB}{\boldsymbol{B}}
\newcommand{\bA}{\boldsymbol{A}}

\newcommand{\feyn}[1]{
  \setbox0=\hbox{\ensuremath{#1}}
  \hbox to\wd0{\hbox to0pt{\hbox to\wd0{\hss/\hss}\hss}\box0}}
\begin{document}

\title{Chiral pumping effect induced by rotating electric fields}
\author{Shu Ebihara}
\affiliation{Department of Physics, The University of Tokyo,
             7-3-1 Hongo, Bunkyo-ku, Tokyo 113-0033, Japan}

\author{Kenji Fukushima}
\affiliation{Department of Physics, The University of Tokyo,
             7-3-1 Hongo, Bunkyo-ku, Tokyo 113-0033, Japan}

\author{Takashi Oka}
\affiliation{Max-Planck-Institut f\"{u}r Physik komplexer Systeme
             (MPI-PKS), N\"{o}thnitzer Stra\ss e 38,
             Dresden 01187, Germany, Max-Planck-Institut f\"{u}r
             Chemische Physik fester Stoffe
             (MPI-CPfS), N\"{o}thnitzer Stra\ss e 40, Dresden 01187,
             Germany}

\begin{abstract}
  We propose an experimental setup using 3D Dirac semimetals to access
  a novel phenomenon induced by the chiral anomaly.  We show that the
  combination of a magnetic field and a circularly polarized laser
  induces a finite charge density with an accompanying axial current.
  This is because the circularly polarized laser breaks time-reversal
  symmetry and the Dirac point splits into two Weyl points, which
  results in an axial-vector field.  We elucidate the appearance of
  the axial-vector field with the help of the Floquet theory by
  deriving an effective Hamiltonian for high-frequency electric
  fields.  This anomalous charge density, i.e.\ the chiral pumping
  effect, is a phenomenon reminiscent of the chiral magnetic effect
  with a chiral chemical potential.  We explicitly compute the pumped
  density and the axial-current expectation value.  We also take
  account of coupling to the chiral magnetic effect to calculate a
  balanced distribution of charge and chirality in a material that
  behaves as a chiral battery.
\end{abstract}
%\pacs{}
\maketitle

%%%%%%%%%%   Introduction   %%%%%%%%%%
\section{Introduction}

Quantum anomaly~\cite{Adler:1969,BellJackiw:1969} is one of the most
fundamental concepts in quantum field theory and anomaly-related
phenomena appear in various physical systems from elementary particle
reactions at high energy to tabletop experiments in condensed matter
physics.  Thus, it would offer tremendous opportunities in theory and
experiment to seek for novel manifestation of quantum anomaly.

Recently, the chiral magnetic effect (CME) has been attracting broad
interest, which was first triggered in the research field of the
ultra-relativistic heavy-ion collision~\cite{Kharzeev:2007jp}.  Later
on, the CME theory has been formulated in terms of the chiral (axial)
chemical potential $\mu_{\rm A}$~\cite{Fukushima:2008xe} and in this
way the application opportunities of the CME have opened to a wider
range of the research fields.  Although it is still challenging to
confirm any smoking-gun experiment through the charge asymmetry
fluctuations in the heavy-ion collision~\cite{Abelev:2009ac},
materials in condensed matter systems would provide us with more
controllable environments.  Recently, the material realization of 3D
Dirac fermions~\cite{Wang,Wang2,Neupane:2014,Liu} and Weyl
fermions~\cite{Cao:2015,Yang:2015,Lv:2015,Xu:2015} has been
established.
In fact, it is claimed that an evidence for the CME has been observed
in ${\rm Zr\,Te_5}$ with parallel electric and magnetic fields, for
which the magnetoconductance has quadratic dependence on the magnetic
field~\cite{Li:2014bha}.

In the present paper, we propose another manifestation of quantum
anomaly which we call the ``chiral pumping effect'' (CPE).  This
effect can be understood as a cousin of the CME originating from a
variant of anomaly relation.  First, let us recall that the CME
relation~\cite{Fukushima:2008xe} is given by a compact formula:
\begin{equation}
  j^z = \frac{e^2 \mu_{\rm A}}{2\pi^2}B^z
  \;\propto\; \epsilon^{3012}A_5^0 F^{12} \quad \mbox{(CME)}\;,
\label{eq:anomalyCME}
\end{equation}
where $\epsilon^{\mu\nu\rho\sigma}$ represents the Levi-Civita
symbol.  This relation states that a charge current $j^z$ is
induced in parallel to the applied magnetic field $B^z=F^{12}$,
provided that the chiral chemical potential $A_5^0$ (= $\mu_{\rm A}$)
is nonzero.  We can anticipate the CPE by exchanging the $3$ and $0$
indices in the above relation, which leads to the following relation
for a charge density:
\begin{equation}
  j^0 \;\propto\; \bA_5\cdot \bB  \quad \mbox{(CPE)}\;.
\label{eq:anomalyCPE}
 \end{equation}
Of course, the net charge must be conserved, so we must understand
this relation in an open system setup.  This means that the charge is
pumped into the system from the surrounding reservoir, and the change
is proportional to the inner product of the axial-vector field $\bA_5$
and the magnetic field.  In the language of Weyl semimetals, $\bA_5$
corresponds to the displacement of two Weyl points.  Since the CPE
originates from quantum anomaly similarly to the CME, we expect that
the coefficient in the relation~\eqref{eq:anomalyCPE} should be
renormalization free.

Here, let us mention on a related approach with an inhomogeneous axion
term $\theta(\bx,t)\bE\cdot\bB$~\cite{Vazifeh:2013,ZWang}.  The
Chern-Simons-Maxwell equations suggest that a charge density,
$j_0\propto (\bnabla\theta)\cdot\bB$, is
induced~\cite{Wilczek:1987mv}.  On the algebraic level this is almost
equivalent to Eq.~\eqref{eq:anomalyCPE} once $\bnabla\theta$ is
identified with $\bA_5$ (which is possible only in the massless case).
The crucial difference in physics is that $\theta(\bx,t)$ should
usually belong to the property of the material, whereas our $\bA_5$ in
Eq.~\eqref{eq:anomalyCPE} is an external field that is experimentally
adjustable as we see later.  It is actually quite non-trivial how to
implement such axial-vector field using electromagnetic devices.  In
the case of experimental confirmation of the CME, in fact, it is
technically difficult to control $\mu_{\rm A}$ and so the
formula~\eqref{eq:anomalyCME} cannot work as it appears for the
experimental signature.  It should be noted that
$\theta(t)=\mu_{\rm A}t$ in the Chern-Simons-Maxwell theory would
immediately lead to Eq.~\eqref{eq:anomalyCME} as pointed out
originally in the very first CME paper~\cite{Fukushima:2008xe} and
later in the context of Weyl semimetals also~\cite{Zyuzin}.  In
contrast, our case of the CPE has an advantage that we can easily
manipulate $\bA_5$.  Moreover, the balanced configuration of charge
and axial-charge (i.e.\ chirality) turns out to be a system of
capacitor of chirality which should be useful for more direct CME
studies.

\begin{figure}
  \includegraphics[width=0.5\columnwidth]{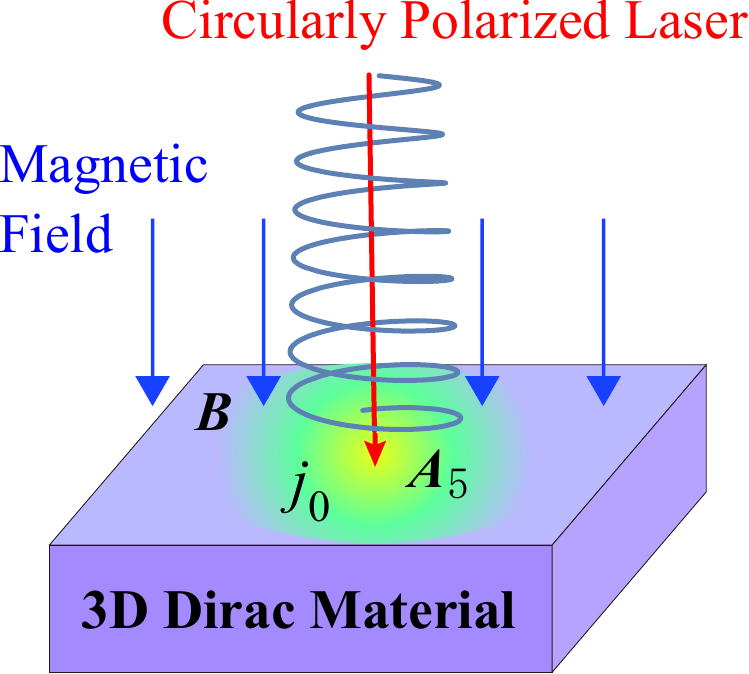}
  \caption{Experimental setup with the magnetic field and the
    circularly polarized laser onto a 3D Dirac semimetal.}
  \label{fig:schematic}
\end{figure}

The aim of this work is to propose a tractable experimental setup to
manifest the CPE in 3D Dirac systems.  A key step to realize the
axial-vector field $\bA_5$ experimentally is, as discussed below, that
we utilize a rotating electric field, i.e., circularly polarized laser
rotating in a two dimensional plane
(see Fig.~\ref{fig:schematic} for a schematic illustration).  We also
refer to a related idea with circular polarizations in 3D Dirac
semimetals~\cite{Narayan} and more general photo-induced
effects~\cite{Chan}.  Using a simple fermionic description, we will
show that the Dirac point splits into two Weyl points.  With an
additional magnetic field~\cite{Nielsen:1983}, a finite density arises
from the lowest Landau level (LLL) of one chirality, which manifests a
concrete picture of the CPE in (1+1)-dimensionally reduced theory of
fermions~\cite{Schon:2000qy}.

This paper is organized as follows.  In Sec.~\ref{sec:Floquet} we
discuss the Floquet effective Hamiltonian to confirm an axial-vector
field.  In Sec.~\ref{sec:response} we consider a combination with a
magnetic field and perform explicit calculations for the charge
density and the axial current.  Inhomogeneous electric charge and
chirality should be balanced with each other.  We solve these coupled
equations of the CPE and the CME to obtain a balanced distribution of
the electric charge and the chirality in Sec.~\ref{sec:distribution}.
Finally, Sec.~\ref{sec:discussions} is devoted to our discussions and
conclusions.

%---   figure   ---%
%---   figure   ---%

%%%%%%%%%%   Effective Hamiltonian   %%%%%%%%%%
\section{Floquet effective Hamiltonian and axial-vector field}
\label{sec:Floquet}

We explain how to realize the axial-vector field in a 3D gapless Dirac
system by applying a circularly polarized laser.  We note that
concrete calculations below are known ones, but a clear recognition of
the axial-vector field has not been established.  When continuous
laser fields are imposed externally, the Hamiltonian $H(t)$ becomes
periodically time-dependent, i.e., $H(t+T)=H(t)$ where $T=2\pi/\Omega$
is the periodicity.  Quantum states in time periodic driving are
described by the Floquet theory~\cite{Sambe:1973,Shirley:1965}, that
is, a temporal version of the Bloch theorem.  The essence of the
Floquet theory is a mapping between the time-dependent Schr\"odinger
equation and a static eigenvalue problem.  The eigenvalue is called
the Floquet pseudo-energy and plays a role similar to the energy in a
static system.  Applications of the Floquet theory to periodically
driven systems with topology changing has been a recent hot
topic~\cite{Oka09,Kitagawa11,Lindner11,Wang:2014} and experiments have
also been done~\cite{Karch10,Karch11,Wang13,Jotzu14}.

To make this paper as self-contained as possible, in this section, we
derive the effective Hamiltonian in an explicit way, though the final
result is not very new but already known.  Let us consider a
Hamiltonian, $H_{\rm tot}=H_0+H_{\rm int}$, with
\begin{equation}
  H_0 = \gamma^0\bgamma\cdot\bp + \gamma^0 m\;, \qquad
  H_{\rm int} = -e\gamma^0\bgamma\cdot\bA\;,
\label{eq:Ham0}
\end{equation}
that describes the one-particle Dirac system coupled to an external
gauge field and $\gamma^\mu$ are the Dirac matrices satisfying
$\{\gamma^\mu,\gamma^\nu\}=2\eta^{\mu\nu}$.  In an electric field with
circular polarization in the $x$-$y$ plane, we can write the
time-dependent vector potentials down as
\begin{equation}
  A_x = \frac{E_0}{\Omega}\cos(\Omega t)\;,\quad
  A_y = \frac{E_0}{\Omega}\sin(\Omega t)\;\quad
  A_z=0\;,
\end{equation}
where $\Omega$ is the frequency.  We can conveniently decompose the
interaction part of the Hamiltonian into two pieces as
$H_{\rm int}=e^{i\Omega t}H_- + e^{-i\Omega t}H_+$ where
$H_\pm=-(eE/\Omega)\gamma^0\gamma^{\pm}$ with
$\gamma^\pm=\frac{1}{2}(\gamma^x\pm i\gamma^y)$.  Now we assume that
the the period $T=2\pi/\Omega$ of the circular polarization is small
enough as compared to the typical observation timescale.  We can then
expand the theory in terms of $\omega/\Omega$ (with $\omega$ being a
frequency corresponding to some excitation energy).  Taking the
average over $T$ we can readily find the following effective
Hamiltonian~\cite{Haeberlen:1968zz,Blanes:2009,Kitagawa10,Kitagawa11}:
\begin{equation}
  H_{\text{eff}} = \frac{i}{T}\ln\Bigl[ \calT
    e^{-i\int_0^T dt\,H(t)} \Bigr]
  \simeq H_0 + \frac{1}{\Omega}[H_-,H_+] \;,
\end{equation}
to the first order in the expansion.  We can also find the same form
from the Floquet Hamiltonian using the Van Vleck perturbation
theory~\cite{Vleck:1929}.
Interestingly we can express the induced term as
\begin{equation}
 H_{\text{ind}} \equiv \frac{1}{\Omega}[H_-,H_+]
  = -\frac{(eE_0)^2}{\Omega^3}i\gamma^x\gamma^y
  = -\beta \gamma^0 \gamma^z \gamma_5 \;,
\label{eq:Hind}
\end{equation}
where we defined $\beta\equiv(eE)^2/\Omega^3$.  This means that the
circular polarized electric field would induce an axial-vector
background field $\bA_5=\beta\hat{\boldsymbol{z}}$ perpendicular to
the polarization plane.  Essentially the same expression was obtained
in the context of ``Floquet Weyl semimetal'' and the corresponding
Floquet bands were figured out~\cite{Wang:2014}.  The physics is
basically the same as this preceding work~\cite{Wang:2014}, but we use
a different language here and, for completeness, we shall see the
energy dispersion relations in the rest of this section. 

\begin{figure}
  \includegraphics[width=0.7\columnwidth]{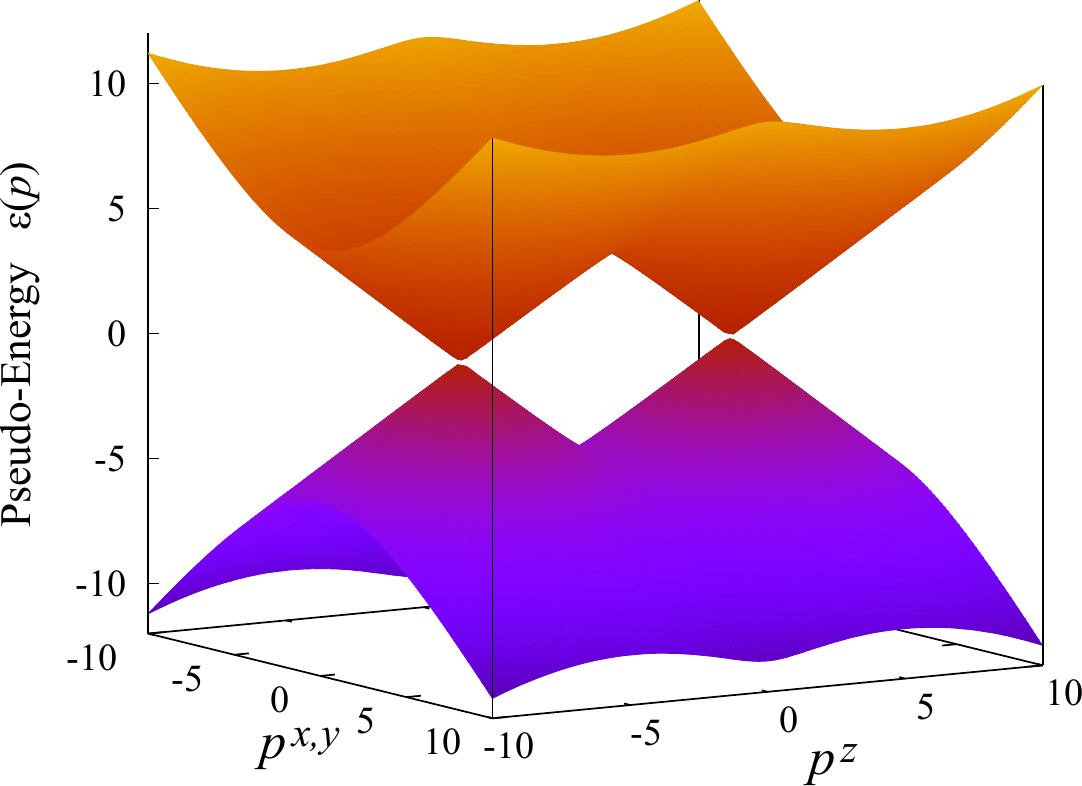}
  \caption{Pseudo-energies $\pm\varepsilon_-(p)$ as a function of
    $p^{x,y}=\sqrt{p_x^2+p_y^2}$ and $p^z$ for $\beta>m$ [in the
      figure $(m,\beta)=(1,5)$ was chosen].}
  \label{fig:disp}
\end{figure}

The effect of finite $\beta$ is easily understandable from the energy
dispersion relations.  We can immediately diagonalize $H_{\text{eff}}$
and the four pseudo-energies read:
\begin{equation}
  \varepsilon_\pm(p)
  = \sqrt{p_x^2 + p_y^2 + (\sqrt{p_z^2+m^2}\pm\beta)^2}
\end{equation}
and $-\varepsilon_\pm(p)$.  We display low-lying $\pm\varepsilon_-(p)$
in Fig.~\ref{fig:disp}. which shows that the Dirac point splits into
two Weyl points with a displacement given by
\begin{equation}
  \Delta p = \sqrt{\beta^2-m^2} \;.
\end{equation}
In fact, $\beta$ is nothing but a momentum shift along the $z$-axis
that is positive for the right chirality state (i.e.,
$\gamma_5\psi_{\rm R}=+\psi_{\rm R}$) and negative for the left
chirality state (i.e., $\gamma_5\psi_{\rm L}=-\psi_{\rm L}$).
We point out that time-resolved ARPES should be able to 
see this splitting of Weyl points in a similar manner as 
the gap opening~\cite{Oka09} of the 2D Dirac point already observed experimentally~\cite{Wang13}.
Interestingly, as long as $\beta>m$, the pseudo-energy always has two
Weyl points (if they are inside of the Brillouin zone) even for $m>0$.
Therefore, we do not have to require strict massless-ness to realize
gapless dispersions, which should be a quite useful feature for
practical applications including the Schwinger or Landau-Zener effect.
In what follows below, we limit ourselves to the $m=0$ case just for
technical simplicity.

%---   figure   ---%
%---   figure   ---%

The generalization from the one-particle Hamiltonian to the many-body
field theory is straightforward.  It is then more convenient to work
with the Lagrangian density corresponding to $H_{\text{eff}}$, that we
can express as
\begin{equation}
  \calL_{\text{eff}} = \bar{\psi}(\feyn{p}-m)\psi
  + \beta\bar{\psi}\gamma^z \gamma_5 \psi 
  + \mu_{\rm A}\bar{\psi}\gamma^0 \gamma_5 \psi\;.
\end{equation}
Here we include $\mu_{\rm A}$ for completeness, which is a necessary
ingredient for the CME.\ \ It is clear from this Hamiltonian that we
should identify $\beta$ as a parameter representing what is called the
\textit{chiral shift}~\cite{Gorbar:2009bm,Gorbar:2010kc}.  
We should emphasize a crucial difference from the idea of the chiral
shift that is not directly controllable but secondarily generated by
finite-density effects.  In our present problem, however, $\beta$ is
an external parameter that we can control with the amplitude and/or
the frequency of the circular polarized electric field.  What we will
see is that, conversely to discussions on the chiral
shift~\cite{Gorbar:2009bm,Gorbar:2010kc}, a finite density is
generated by this externally given $\beta\neq0$.

%%%%%%%%%%   Response to the magnetic field   %%%%%%%%%%
\section{Response to the magnetic field}
\label{sec:response}

%---   figure   ---%
\begin{figure}
  \includegraphics[width=0.95\columnwidth]{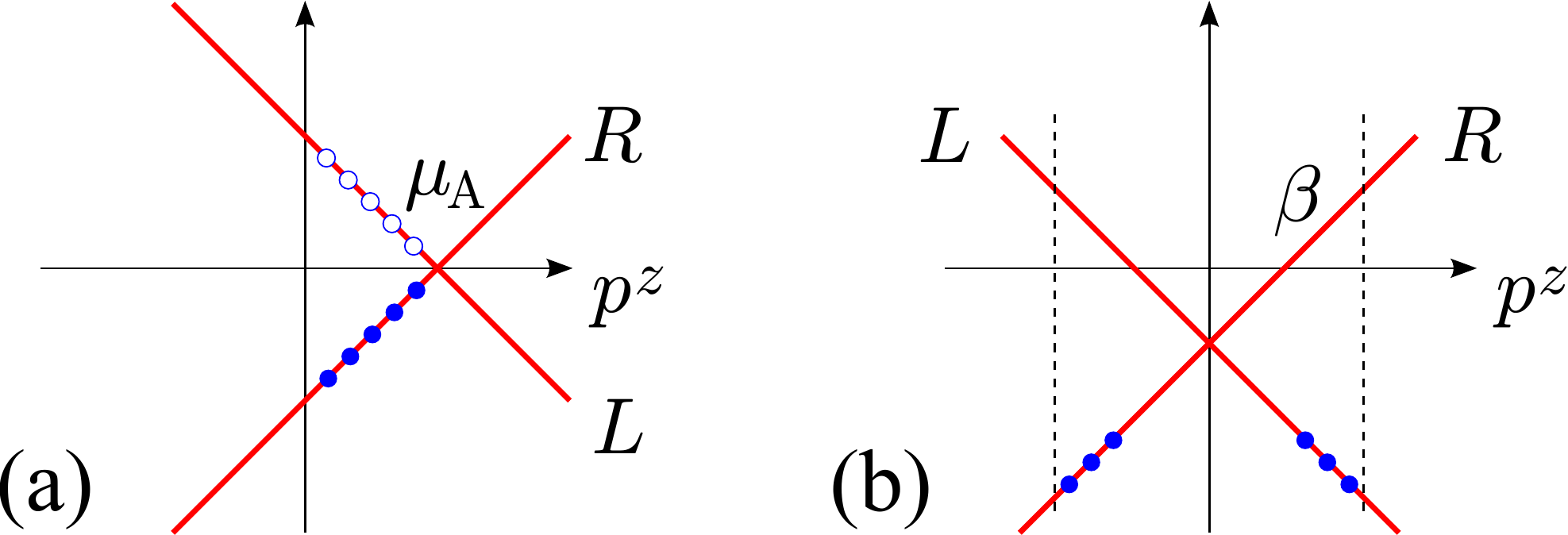}
  \caption{Energy dispersion relations of the LLL (i.e., $p^x=p^y=0$)
    with (a) chiral chemical potential $\mu_{\rm A}$ relevant for the
    CME and (b) chiral shift $\beta$ relevant for the CPE.\ \
    Filled (blank) dots represent states that are newly occupied
    (unoccupied).}
  \label{fig:shift}
\end{figure}
%---   figure   ---%

It is the most essential point that we can regard $\beta$ as the $z$
component of an axial-vector field; $\beta\sim A_5^z$.  Then,
if we further impose an external magnetic field $B^z=F^{12}$ on this
system, we should expect the following anomaly relation;
%\begin{equation}
$j^0 \;\propto\; \beta B$,
%\label{eq:anomaly}
%\end{equation}
immediately from the CPE or Eq.~\eqref{eq:anomalyCPE}.  We will
confirm this expectation with explicit calculations, but before going
into details, let us consider an intuitive interpretation to
understand Eq.~\eqref{eq:anomalyCPE}.

For the purpose of comprehensible illustration it would be useful to
sketch the dispersion relations in the same way as in a CME
literature~\cite{Fukushima:2008xe}.
Figure~\ref{fig:shift}~(a) shows the dispersion relations of the
lowest Landau levels (LLLs) relevant for the CME with $\beta=0$ and
$\mu_{\rm A}\neq0$.  In this case with $\mu_{\rm A}>0$ the energies of
the right-handed ($R$) particles are decreased, while those of the
left-handed ($L$) particles are increased.  Note that only one spin
state is chosen out for the LLL depending on the sign of $eB$.
Therefore, a positive $\mu_{\rm A}$ favors more $R$ than $L$.  This
explains how a finite chiral density is accumulated, while a net
density remains vanishing.  Also we see that the CME current flows
from the LLL with $\mu_{\rm A}$ and gives an intuitive picture for the
anomaly relation~\eqref{eq:anomalyCME}.  In contrast, as seen in
Fig.~\ref{fig:shift}~(b) for $\beta>0$, the energy dispersion of $R$
(and $L$) is shifted positively (and negatively, respectively) along
the $p^z$ axis.  Thus, assuming that particles can flow in through the
$p^z$-integration edges [as indicated by the dashed lines in
Fig.~\ref{fig:shift}~(b)], both $L$ and $R$ LLL states increase their
occupation.   This is the mechanism of how a finite density develops
from a combination of $\beta$ and $B$ (that causes the dimensional
reduction) through the anomaly relation~\eqref{eq:anomalyCPE}.  The
argument at the same time tells us that a $R$ excess in $p_z<0$ and an
$L$ excess in $p_z>0$ contribute to a negative flow of the net
chirality, which is an analogue of the CME current.  Summarizing the
above discussions we should expect non-zero $\langle j^0\rangle$ and
$\langle j_{\rm A}^z\rangle$.

Now we shall make sure of our qualitative argument by performing
explicit calculations and identify the coefficient of
Eq.~\eqref{eq:anomalyCPE} following the calculations done
previously~\cite{Gorbar:2009bm,Gorbar:2010kc}.  We set $\mu_{\rm A}=0$
in the following and concentrate on the CPE only.  The fermion
propagator with $\beta$ in the presence of $eB>0$ is
\begin{equation}
  G(x\sim 0) \simeq G_0^-(x) P_-
  + \sum_{n=1}^\infty [G_n^+(x) P_+ + G_n^-(x) P_-]
\end{equation}
near the coincidence limit, where
$P_\pm\equiv\frac{1}{2}(1\pm i\gamma^x\gamma^y)$ is the spin
projection operator and the first term, $G_0^- P_-$, is the LLL
contribution.  The (Feynman) propagator for each Landau mode is
\begin{align}
   G_n^\pm(x\sim 0) &= \frac{ieB\gamma^0}{2\pi}\int\frac{d\omega\, dp}
   {(2\pi)^2} \Bigl[ K_n(\omega,\pm(p+\beta))P_5^- \notag\\
   &\qquad\qquad + K_n(\omega,\mp(p-\beta))P_5^+ \Bigr] \;.
\end{align}
Here, $P_5^\pm\equiv\frac{1}{2}(1\pm\gamma_5)$ is the chirality
projection operator and we defined
\begin{equation}
  K_n(\omega,p) = \frac{\omega+p}{\omega^2-2eB n-p^2+i\epsilon}\;.
\end{equation}
Using this form of the propagator we can express the current
expectation value as
\begin{equation}
  \langle j^0\rangle = -e\,\lim_{t\to0^+}\tr[\gamma^0 G(t)]\;.
\end{equation}
We can simplify the calculation by noting that
$\tr(P_5^\pm P_\pm)=\frac{1}{4}\tr[(1\pm\gamma_5)(1\pm i\gamma^x\gamma^y)]
=1$ where two $\pm$ are independent.  Because $\omega$ in the
numerator should be vanishing after the $\omega$ integration, it is
straightforward to check that
$\tr(\gamma^0G_n^+P_+)+\tr(\gamma^0G_n^-P_-)=0$.  Therefore only the
LLL contribution survives, which yields:
\begin{equation}
  \langle j^0\rangle = -\frac{e^2B}{2(2\pi)} \int\frac{dp}{2\pi}
  \biggl( -\frac{p+\beta}{|p+\beta|} + \frac{p-\beta}{|p-\beta|}
  \biggr)
\end{equation}
after the $\omega$ integration.  The above is a finite integral only
from $-\beta\le p_z\le\beta$.  Actually, this simple expression is
precisely a concrete realization of our qualitative argument
based on Fig.~\ref{fig:shift}~(b).  Finally, after the $p$
integration, we recover the same result as previously obtained
one~\cite{Gorbar:2009bm}:
\begin{equation}
  \langle j^0\rangle = \frac{e^2 \beta B}{2\pi^2} \;,
\label{eq:j0}
\end{equation}
which is quite reminiscent of the formula for the CME current, namely,
$\langle j^z\rangle=e^2 \mu_{\rm A} B/(2\pi^2)$.  This compact formula
itself is a known one, but the fact that the above
$\langle j^0\rangle$ can be easily realized with a circularly
polarized laser is our main claim in this work.

Remembering our qualitative argument, we can in turn expect an axial
current $\langle j_{\rm A}^z\rangle$ which is similar to what is
called the chiral separation effect;
$\langle j_{\rm A}^z\rangle \propto \mu B$~\cite{Metlitski:2005pr}.
To see this, let us perform an explicit calculation using
$\langle j_{\rm A}^z\rangle=-e\,\tr[\gamma^z\gamma_5 G(0^+)]$.  In
this case we use $\tr(\gamma^z\gamma_5\gamma^0 P_5^\pm P_\pm)
=\frac{1}{4}\tr[\gamma^z\gamma_5\gamma^0(1\pm\gamma_5)
  (1\pm i\gamma^x\gamma^y)]
=\pm\frac{i}{4}\tr(\gamma^z\gamma_5\gamma^0\gamma^x\gamma^y)=\pm 1$,
where the last $\pm$ refers to $\pm$ of $P_\pm$.  Then, the LLL
contribution is completely identical to that for $\langle j^0\rangle$
except for the overall sign, that means,
$\langle j_{\rm A}^z\rangle_{\text{(LLL)}}=-\langle j^0\rangle$. 
We can immediately explain this proportionality for the LLL from the
property of the two-dimensional Dirac matrices;
$\gamma^z\gamma_5^{(2)}=\gamma^0$ where $\gamma_5^{(2)}$ is the
two-dimensional counterpart defined by $\gamma^0\gamma^z$.  In this
case, however, some complication appears from non-zero Landau levels
as is the case for the chiral
shift~\cite{Gorbar:2009bm,Gorbar:2010kc}.  We can write those
contributions down explicitly as
\begin{equation}
  \begin{split}
  &\langle j_{\rm A}^z\rangle_{(n>0)} = \frac{e^2B}{2\pi}\int
  \frac{dp}{2\pi}\sum_n \\
  & \times \biggl( -\frac{p+\beta}
       {\sqrt{2eB n+(p\!+\!\beta)^2}} + \frac{p-\beta}
       {\sqrt{2eB n+(p\!-\!\beta)^2}} \biggr) \;.
  \end{split}
\end{equation}
This is a subtle expression whose precise value depends on how to
organize the infinity in the Landau sum and the momentum integration.
For example, if we take the Landau sum up to $N$ and the momentum
integration within $|p|<\Lambda$ in such a way that
$2eBN\ll\Lambda^2$, then the $p$ integration is easy to perform and we
find $\langle j_{\rm A}^z\rangle_{(n>0)}=-2N\langle j^0\rangle$.  For
more general situation $N$ should be replaced with
$(\Lambda/eB)(\sqrt{2eBN+\Lambda^2}-\Lambda)$, which approaches
$\sqrt{2N\Lambda^2/eB}$ in the opposite limit of $\Lambda^2\ll 2eBN$.
In reality these cutoffs should be fixed by the microscopic properties
of the material, especially the Debye mass and the Brillouin zone
structures.  The conclusion is that the system comes to have a finite
axial-current as given  by
\begin{equation}
  \langle j_{\rm A}^z\rangle = -\alpha\,\frac{e^2 \beta B}{2\pi^2} \;,
\label{eq:jA}
\end{equation}
the coefficient $\alpha$ of which is not anomaly protected unlike
Eq.~\eqref{eq:j0} and it is a material-dependent problem to fix
$\alpha$.

%%%%%%%%%%   Static charge distribution   %%%%%%%%%%
\section{Static charge distribution}
\label{sec:distribution}

So far, we assumed an infinitely large system.  Imposing surface
boundary conditions with a finite extent, we should take account of
the polarization and screening effects to find a balanced distribution
of the charge and the chirality density.  Let us consider a
finite-size material whose thickness in the $z$ direction is $d$;  we
take the $z$ coordinate so that the material is placed in a range
$-d/2 \leq z \leq +d/2$.  Then, because of the surface effects, we
should introduce $z$ dependent $\mu(z)$ and $\mu_{\rm A}(z)$, the
determination of which is the goal of this section.

In the presence of $\mu_{\rm A}(z)$, the CME current should follow
from Eq.~\eqref{eq:anomalyCME}.  In equilibrium, however, there should
be no current and the CME current should be canceled by the
polarization effect, i.e.
\begin{equation}
 j^z_{\rm total} = \frac{e^2 B}{2\pi^2}\mu_{\rm A}(z)
  + \sigma \int dz\, j^0(z) \;,
\label{eq:balance_j}
\end{equation}
where the second term in the right-hand side appears from Ohm's law
with the electric conductivity $\sigma>0$.  In fact, $\int dz\,j^0(z)$
is nothing but an electric field associated with the polarization or
the charge distribution $j^0(z)$.  Qualitatively, this last term
represents a flow to flatten the charge distribution.  For the balance
equation for $j_{\rm A}^z$ we postulate the same structure as
Eq.~\eqref{eq:balance_j} with the replacement of
$j^\mu \leftrightarrow j_{\rm A}^\mu$ and $\mu\leftrightarrow \mu_{\rm A}$
(i.e., the chiral separation effect~\cite{Metlitski:2005pr}):
\begin{equation}
  j_{\rm A\,,total}^z = \frac{e^2B}{2\pi^2}\mu(z) 
               - \frac{e^2 B}{2 \pi^2}\alpha \beta 
%                         + \sigma_A \int dz j_0^A \;.
               - \lambda \partial_z j_{\rm A}^0(z) \;,
\end{equation}
where we added the last term to consider diffusion processes of
chirality introducing a diffusion constant $\lambda>0$.  Intuitively,
this last term represents movement of chirality to decrease the
chirality gradient.  The densities, $j^0$ and $j_{\rm A}^0$, are given
as
\begin{equation}
 j^0 = \frac{e^2 B}{2 \pi^2} \beta + \frac{e}{3 \pi^2} \mu(z)^3 \;,
 \qquad
 j_{\rm A}^0 = \frac{e}{3 \pi^2} \mu_{\rm A}(z)^3 \;,
\end{equation}
where a CPE contribution to $j^0$ is added to the standard
density-chemical potential relation.  We plug these expressions to the
conservation laws; $\partial_{\mu} j^{\mu} = 0$ and
$\partial_{\mu} j_{\rm A}^{\mu} = 0$ (note that there is no
$\bE\cdot\bB$ with our electromagnetic configuration).  Since we are
interested in the static profile only, we can drop the
time-derivatives to reach finally:
\begin{align}
 & \frac{e^2B}{2\pi^2} \partial_z \mu_{\rm A}(z) 
   + \sigma \frac{e^2B}{2\pi^2} \beta 
   + \frac{e\sigma}{3 \pi^2} \mu(z)^3 = 0 \;,
\label{eq:balance} \\
 & \frac{e^2B}{2\pi^2} \partial_z \mu(z)
   - \frac{e\lambda}{3 \pi^2} \partial_z^2 \mu_{\rm A}(z)^3 = 0 \;.
\end{align}
We should solve these differential equations under the charge
constraint; $\int_{-d/2}^{d/2} dz\, j^0(z) = (e^2 B/2\pi^2)\beta d$,
which leads to $\mu_{\rm A}(d/2)-\mu_{\rm A}(-d/2)=-\beta\sigma d$.
If the thickness $d$ of the material is
sufficiently small, we can find an approximate solution by expanding
the solutions in terms of $z$, that is,
\begin{equation}
 \mu(z) \simeq -\frac{2\beta^3 \sigma^3\lambda}{eB} z^2\;,
 \qquad
 \mu_{\rm A}(z) \simeq -\beta\sigma z\;.
\end{equation}
Here, we further used a condition to impose zero net chirality.  It
should noted that these are leading-order results and, to satisfy
Eq.~\eqref{eq:balance} strictly for example, we should consider a
term $\propto z^7$ in $\mu_{\rm A}(z)$ which we neglected.  The
charge density is thus more screened near the surface at larger $|z|$
unlike an ordinary conductor, while there are more $R$ (and $L$) near
the bottom at negatively large $z$ (and near the top at positively
large $z$, respectively).  This is very interesting result; the system
behaves as a capacitor of chirality or the chiral battery in which a
non-zero slope in the chirality distribution is sustained by the CPE
as well as a finite net density.  Such a realization of the chiral
battery should be useful for more quantitative investigations of the
CME and related phenomena, e.g., the chiral plasma
instability~\cite{Akamatsu:2013pjd} for instance.  We note that the
idea of the chiral battery can be traced back to the original
work~\cite{Fukushima:2008xe} and extensively discussed as chiral
electronics~\cite{Kharzeev:2013}.

%%%%%%%%%%   Discussions   %%%%%%%%%%
\section{Discussions and summary}
\label{sec:discussions}

The anomalously induced charge density $\langle j^0\rangle$  given by
Eq.~\eqref{eq:anomalyCPE} or \eqref{eq:j0} is an experimentally
detectable quantity.  One possible way to observe this is using
transient grating techniques~\cite{Gedik:2003}.  The grating of the
circular polarized field would lead to inhomogeneous charge
distribution, and the dynamics afterward can be measured.

It is also an interesting possibility to make use of Eq.~\eqref{eq:jA}
instead of Eq.~\eqref{eq:j0} since $\alpha$ could take a large number
(involving some cutoffs at least in a naive estimate).  In the
two-component spinor representation,
$j_{\rm A}^z=\phi^\ast_{\rm R} \sigma^z\phi_{\rm R}
+ \phi^\ast_{\rm L} \sigma^z\phi_{\rm L}$ is nothing but the spin
expectation value.  Hence, for systems with gapless Dirac particles,
Eq.~\eqref{eq:jA} indicates not only the dynamical flow of chirality
but also the spin polarization.  Spin polarization can be probed by
pump-probe magneto-optical Kerr effect.  In addition, when topological
current or density exists in general, there can be some photo-emission
processes via anomaly as discussed in
Refs.~\cite{Basar:2012bp,Fukushima:2012fg}.  It deserves further
investigations to quantify emitted photon spectra associated with the
CME and the CPE.\ \ We will leave them for interesting future
problems.

In summary, we formulated the CPE that generates a finite density from
the combination of the circular polarized electric field and the
magnetic field.  The interesting point in this setup is that the
externally imposed electric and magnetic fields are always pointing
orthogonal to each other and there is no inner product of
$\bE\cdot\bB$ which is usually the source of topological charge and
parity breaking.  Instead, the circular polarization breaks the parity
symmetry, and so it alone can take care of the role of $\bE\cdot\bB$.
Besides, thanks to rotating electric field, we do not need to subtract
a huge background due to ordinary electric current that obeys Ohm's
law.

We gave intuitive arguments about the generation of the density and
the axial current and confirmed our expectations by explicit
calculations recovering previously known expressions.  We saw that
the density $\langle j^0\rangle$ is expressed in a compact form
similar to the CME and its coefficient is anomaly protected.  In
contrast to this, $\langle j_{\rm A}^z\rangle$ has complicated
contributions that need some ultraviolet regularization or physical
cutoff.  We would here emphasize the following point:  Each building
block for our conclusion was known;  the appearance of a
$\gamma^x\gamma^y$ term from the Floquet Weyl semimetal and the
topologically induced density from the chiral shift were all
``re-derived'' in this work, but our novelty is found in a combination
of them.  The most important is that our controlling parameter $\beta$
is given externally unlike the axion term associated with intrinsic
properties of the material.

Although it is difficult to determine the coefficient of
$\langle j_{\rm A}^z\rangle$, it is still proportional to $\beta$ or a
combination of $(eE)^2/\Omega^3$ (as long as $\Omega$ is large enough
to justify the leading-order expansion as we did in this work).  This
characteristic dependence of $\Omega^{-3}$ should give a consistency
check for anomalously induced $\langle j^0\rangle$ and
$\langle j_{\rm A}^z\rangle$.

Fortunately, such ambiguity in $\langle j_{\rm A}^z\rangle$ does not
enter the determination of the static charge and chirality
distribution in the material.  Our study implies that the induced
$\langle j^0\rangle$ gets slightly smaller near the material surfaces.
The most interesting is that the chirality has a non-trivial
distribution also and the chirality separation is sustained by the
CPE, which realizes a system of the chiral battery.  The experimental
confirmation of the CPE itself is quite challenging, and furthermore,
the CPE opens a new possibility for more direct CME studies using this
chiral battery.

\acknowledgments
K.~F.\ thanks Dima~Kharzeev for useful discussions.
We also thank Eduard~Gorbar, Volodya~Miransky, and Igor~Shovkovy for
encouraging comments.
K.~F.\ was partially supported by JSPS KAKENHI Grant
No.\ 15H03652 and 15K13479,
and T.~O.\ by No.\ 26400350.


\begin{thebibliography}{99}

\bibitem{Adler:1969} 
 S.~Adler, 
  Phys.\ Rev. {\bf 177}, 2426 (1969).
  
\bibitem{BellJackiw:1969} 
 J.~S.~Bell, and R.~Jackiw,
  Nuovo Cimento {\bf 60A}, 4 (1969).

\bibitem{Kharzeev:2007jp} 
  D.~E.~Kharzeev, L.~D.~McLerran and H.~J.~Warringa,
  %``The Effects of topological charge change in heavy ion collisions: 'Event by event P and CP violation',''
  Nucl.\ Phys.\ A {\bf 803}, 227 (2008).
%  [arXiv:0711.0950 [hep-ph]].

\bibitem{Fukushima:2008xe} 
  K.~Fukushima, D.~E.~Kharzeev and H.~J.~Warringa,
  %``The Chiral Magnetic Effect,''
  Phys.\ Rev.\ D {\bf 78}, 074033 (2008).
%  [arXiv:0808.3382 [hep-ph]].

\bibitem{Abelev:2009ac}
  B.~I.~Abelev {\it et al.} [STAR Collaboration],
  %``Azimuthal Charged-Particle Correlations and Possible Local Strong Parity Violation,''
  Phys.\ Rev.\ Lett.\  {\bf 103}, 251601 (2009);
%  [arXiv:0909.1739 [nucl-ex]].
%\bibitem{Abelev:2012pa} 
  B.~Abelev {\it et al.} [ALICE Collaboration],
  %``Charge separation relative to the reaction plane in Pb-Pb collisions at $\sqrt{s_{NN}}= 2.76$ TeV,''
  Phys.\ Rev.\ Lett.\  {\bf 110}, 012301 (2013).
%  [arXiv:1207.0900 [nucl-ex]].


\bibitem{Wang}
 Z.~Wang {\it et al.},
 Phys.\ Rev.\ B {\bf 85}, 195320 (2012).

\bibitem{Wang2}
 Z.~Wang, H.~Weng, Q.~Wu, X.~Dai, and Z.~Fang,
 Phys.\ Rev.\ B {\bf 88}, 125427 (2013).

\bibitem{Neupane:2014} 
M.~Neupane {\it et al.},
  %``Observation of a topological 3D Dirac semimetal phase in high-mobility Cd 3 As 2,''
  Nat.\ Phys.\ {\bf 5},  4086 (2014) 

\bibitem{Liu}
 Z.K.~Liu {\it et al.},
 Nat.\ Mat.\ {\bf 13}, 677--681 (2014).

\bibitem{Cao:2015} 
  J.~Cao {\it et al.},
  %``Landau level splitting in Cd3As2 under high magnetic fields,''
  Nat.\ Com.\ {\bf 6}, 7779 (2015) 

\bibitem{Yang:2015} 
L.~X.~Yang {\it et al.},
  %``Weyl semimetal phase in the non-centrosymmetric compound TaAs,''
  Nat.\ Phys.\ {\bf 10}, 3425 (2015) 

\bibitem{Lv:2015} 
  B.~Lv {\it et al.},
  %``Observation of Weyl nodes in TaAs,''
  Nat.\ Phys.\ {\bf 10}, 3426 (2015) 

\bibitem{Xu:2015} 
S.~Y.~Xu {\it et al.},
  %``Discovery of a Weyl fermion state with Fermi arcs in niobium arsenide,''
  Nat.\ Phys.\ {\bf 10}, 3437 (2015) 
  

\bibitem{Li:2014bha} 
  Q.~Li {\it et al.},
  %``Observation of the chiral magnetic effect in ZrTe5,''
  arXiv:1412.6543 [cond-mat.str-el].


\bibitem{Vazifeh:2013}
  M.M.~Vazifeh and M.~Franz,
  Phys.\ Rev.\ Lett.\ {\bf 111}, 027201 (2013).

\bibitem{ZWang}
 Z.~Wang and S.-C.~Zhang,
 Phys.\ Rev.\ B {\bf 87}, 161107 (2013).

\bibitem{Wilczek:1987mv} 
  F.~Wilczek,
  %``Two Applications of Axion Electrodynamics,''
  Phys.\ Rev.\ Lett.\  {\bf 58}, 1799 (1987).
%  doi:10.1103/PhysRevLett.58.1799

\bibitem{Zyuzin}
  A.A.~Zyuzin and A.A.~Burkov,
  Phys.\ Rev.\ B {\bf 86}, 115133 (2012).

\bibitem{Narayan}
  A.~Narayan,
  Phys.\ Rev.\ B {\bf 91}, 205445 (2015).

\bibitem{Chan}
  C.-K.~Chan, P.A.~Lee, K.S.~Burch, J.H.~Han, and Y.~Ran,
  arXiv:1509.05400 [cond-mat.mes-hall].

\bibitem{Nielsen:1983} 
H.~B.~Nielsen, M.~Ninomiya,
  %``The Adler-Bell-Jackiw anomaly and Weyl fermions in a crystal,''
 Phys.\ Lett. \ {\bf 130B},  389 (1983)

\bibitem{Schon:2000qy} 
  V.~Schon and M.~Thies,
  %``2-D model field theories at finite temperature and density,''
  In *Shifman, M. (ed.): At the frontier of particle physics, vol. 3* 1945-2032
  [hep-th/0008175].

\bibitem{Sambe:1973}
H.~Sambe,
  Phys.\ Rev.\ A {\bf 7}, 2203 (1973).

\bibitem{Shirley:1965}
  J.H.~Shirley,
  Phys.\ Rev.\ B {\bf 138}, 979 (1965).

\bibitem{Oka09} % Photo-induced topological phase transition
T. Oka and H. Aoki, 
  Phys.\ Rev.\ B {\bf 79}, 081406 (2009).

\bibitem{Kitagawa11} % Floquet topological insulator
T. Kitagawa, T. Oka, A. Brataas, L. Fu, and E. Demler, 
  Phys.\ Rev.\ B {\bf 84}, 235108 (2011).

\bibitem{Lindner11} % Floquet topological insulator
N. H. Lindner, G. Refael, and V. Galitski, 
  Nat.\ Phys. {\bf 7}, 490 (2011).
  
\bibitem{Wang:2014} 
  R.~Wang, B.~Wang, R.~Shen, L.~Sheng and D. Y.~Xing,
  %``Floquet Weyl semimetal induced by off-resonant light %topological phase transitions,''
 EPL {\bf 105}, 17004 (2014).



\bibitem{Karch10}
J. Karch, {\it et al.} , 
  Phys.\ Rev.\ Lett.\  {\bf 105}, 227402 (2010).
  
\bibitem{Karch11}
J. Karch, {\it et al.} , 
  Phys.\ Rev.\ Lett.\  {\bf 107}, 276601 (2011).

\bibitem{Wang13}  % Observation of Floquet topological insulator
Y. H. Wang, H. Steinberg, P. Jarillo-Herrero, and N. Gedik, 
Science, {\bf 342}, 453 (2013). 

\bibitem{Jotzu14}
G. Jotzu, M. Messer, R\'emi Desbuquois, M. Lebrat, T. Uehlinger, 
D. Greif, and T. Esslinger, Nature {\bf 515}, 237 (2014). 

\bibitem{Kitagawa10} % Floquet effective Hamiltonian
T. Kitagawa, M. S. Rudner, E. Berg, and E. Demler, 
  Phys.\ Rev.\ A {\bf 82}, 0033429 (2010).
\bibitem{Haeberlen:1968zz} 
  U.~Haeberlen and J.~S.~Waugh,
  %``Coherent Averaging Effects in Magnetic Resonance,''
  Phys.\ Rev.\  {\bf 175}, 453 (1968).

\bibitem{Blanes:2009}
  S.~Blanes, F.~Casas, J.A.~Oteo and J.~Ros,
  %``The Magnus expansion and some of its applications,''
  Phys.\ Rept.\ {\bf 470}, 151--238 (2009).

\bibitem{Vleck:1929}
  J.~H.~Van~Vleck,
  Phys.\ Rev.\ {\bf 33}, 467 (1929).

\bibitem{Gorbar:2009bm} 
  E.~V.~Gorbar, V.~A.~Miransky and I.~A.~Shovkovy,
  %``Chiral asymmetry of the Fermi surface in dense relativistic matter in a magnetic field,''
  Phys.\ Rev.\ C {\bf 80}, 032801 (2009);
%  [arXiv:0904.2164 [hep-ph]].
%\bibitem{Gorbar:2011ya} 
%  E.~V.~Gorbar, V.~A.~Miransky and I.~A.~Shovkovy,
  %``Normal ground state of dense relativistic matter in a magnetic field,''
  Phys.\ Rev.\ D {\bf 83}, 085003 (2011).
%  doi:10.1103/PhysRevD.83.085003
%  [arXiv:1101.4954 [hep-ph]].

\bibitem{Gorbar:2010kc} 
  E.~V.~Gorbar, V.~A.~Miransky and I.~A.~Shovkovy,
  %``Chiral asymmetry and axial anomaly in magnetized relativistic matter,''
  Phys.\ Lett.\ B {\bf 695}, 354 (2011).
%  [arXiv:1009.1656 [hep-ph]].

\bibitem{Metlitski:2005pr} 
  M.~A.~Metlitski and A.~R.~Zhitnitsky,
  %``Anomalous axion interactions and topological currents in dense matter,''
  Phys.\ Rev.\ D {\bf 72}, 045011 (2005).
%  [hep-ph/0505072].

\bibitem{Akamatsu:2013pjd} 
  Y.~Akamatsu and N.~Yamamoto,
  %``Chiral Plasma Instabilities,''
  Phys.\ Rev.\ Lett.\  {\bf 111}, 052002 (2013).
%  doi:10.1103/PhysRevLett.111.052002
%  [arXiv:1302.2125 [nucl-th]].
  %%CITATION = doi:10.1103/PhysRevLett.111.052002;%%

\bibitem{Kharzeev:2013}
  D.E.~Kharzeev and H.-U.~Yee,
  Phys.\ Rev.\ B {\bf 88}, 115119 (2013).

\bibitem{Gedik:2003} 
  N.~Gedik, J.~Orenstein, R.~Liang, D.A.~Bonn, W.N.~Hardy,
  %``Diffusion of nonequilibrium quasi-particles in a cuprate superconductor.,''
  Science {\bf 300}, 1410 (2003).
%  [arXiv:1206.3128 [hep-ph]].


\bibitem{Basar:2012bp} 
  G.~Basar, D.~Kharzeev and V.~Skokov,
  %``Conformal anomaly as a source of soft photons in heavy ion collisions,''
  Phys.\ Rev.\ Lett.\  {\bf 109}, 202303 (2012).
%  [arXiv:1206.1334 [hep-ph]].

\bibitem{Fukushima:2012fg} 
  K.~Fukushima and K.~Mameda,
  %``Wess-Zumino-Witten action and photons from the Chiral Magnetic Effect,''
  Phys.\ Rev.\ D {\bf 86}, 071501 (2012).
%  [arXiv:1206.3128 [hep-ph]].




\end{thebibliography}
\end{document}